\documentclass[twocolumn]{webofc}
\usepackage[varg]{txfonts}   % Web of Conferences font
\begin{document}
\title{The Pulsed Neutron Beam EDM Experiment}

\author{
\firstname{Estelle} \lastname{Chanel}\inst{1}
\and
\firstname{Zachary} \lastname{Hodge}\inst{1}
\and
\firstname{Dieter} \lastname{Ries}\inst{1}\fnsep\thanks{New address: Institute of Nuclear Chemistry, Johannes Gutenberg University Mainz, Germany}
\and
\firstname{Ivo} \lastname{Schulthess}\inst{1}
\and
\firstname{Marc} \lastname{Solar}\inst{1}
\and
\firstname{Torsten} \lastname{Soldner}\inst{2}
\and %\\
\firstname{Oliver} \lastname{Stalder}\inst{1}
\and 
\firstname{Jacob} \lastname{Thorne}\inst{1}
\and
\firstname{Florian M.} \lastname{Piegsa}\inst{1}\fnsep\thanks{\email{florian.piegsa@lhep.unibe.ch}}
}

%\author{
%\firstname{F.M.} \lastname{Piegsa}\inst{1}\fnsep\thanks{\email{florian.piegsa@lhep.unibe.ch}} \and
%\firstname{E.} \lastname{Chanel}\inst{1} \and
%\firstname{Third author} \lastname{Third author}\inst{1}\fnsep\thanks{\email{Mail address for last author if necessary}}
%}

\institute{
Laboratory for High Energy Physics, Albert Einstein Center for Fundamental Physics, University of Bern, 3012 Bern, Switzerland
\and
Institut Laue-Langevin, CS 20156, 38042 Grenoble Cedex 9, France
}

\abstract{
We report on the Beam EDM experiment, which aims to employ a pulsed cold neutron beam to search for an electric dipole moment instead of the established use of storable ultracold neutrons. 
We present a brief overview of the basic measurement concept and the current status of our proof-of-principle Ramsey apparatus. % experiment.
}
\maketitle
\section{Introduction}
\label{intro}

The measurement of the neutron electric dipole moment (EDM) is considered to be one of the most important fundamental physics experiments at low energy. It presents a very promising route for finding new physics beyond the standard model of particle physics and violations of the combined charge parity symmetry CP which could explain the observed matter-antimatter baryon asymmetry in the universe. The current limit for a neutron EDM has reached an impressive value of $3 \times 10^{-26}$~e~cm \cite{Pendle15}.
The search for a finite neutron EDM remains a top priority. It has become a worldwide endeavor which is followed by various research teams setting up experiments for improved measurements \cite{Baker11,Abel18,Serebrov15,Ito07,Tsentalovich2014,Lamoreaux09,Altarev12,Picker17,Masuda2012}. All these collaborations are using or are planning to use ultracold neutrons, since they have the advantage of long interaction times. However, historically early neutron EDM experiments have been performed employing neutron beams where the main limiting systematic has so far been the relativistic  $v \times E$-effect.  This effect arises from the neutron moving at a velocity $v$ through an electric field $E$ sensing an effective magnetic field $\vec{B}_{v \times E}  = - (\vec{v} \times \vec{E})/c^2$ according to Maxwell’s equations, with $c$ being the speed of light in vacuum. In the most recent neutron EDM beam experiment, performed in the 1970s, this effect was corrected for by mounting the entire Ramsey spectrometer on a large turntable, in order to reverse the direction of the neutron beam \cite{Dress77}. The final result in this experiment reached a limit of $3 \times 10^{-24}$~e~cm.
Recently, a novel concept has been proposed to measure the neutron EDM with a cold pulsed neutron beam instead of the established use of stored ultracold neutrons \cite{Piegsa13}. The technique relies on the fact that one can distinguish between the effect due to a neutron EDM and the $v \times E$-effect by performing a time-of-flight Ramsey measurement. \\
The new method can ultimately lead to a highly competitive result with different sensitivities to possible systematic effects. Currently, we develop a scaled-down proof-of-principle Ramsey apparatus and perform several experiments at different neutron beam lines at the Paul Scherrer Institute (PSI) and the Institut Laue-Langevin (ILL). The findings of these investigations are essential for a later design of the full-scale experiment, intended for the upcoming European Spallation Source (ESS) in Sweden and its planned fundamental physics beam line ANNI \cite{Klinkby16,Soldner18}.

\section{Ramsey beam technique}

A neutron EDM is measured most commonly by employing Ramsey’s technique of separated oscillatory fields \cite{Ramsey49,Ramsey50}. 
Over the last years, the technique has been applied in various different precision neutron beam experiments \cite{Piegsa08,Piegsa09,Piegsa12,Piegsa14}.
In such a measurement, polarized neutron spins, exposed to an external static magnetic field $B_0$, are irradiated with two consecutive phase-locked radio-frequency $\pi/2$-pulses. A so-called Ramsey oscillation pattern is obtained by scanning the frequency $\omega$ of the pulses close to the Larmor resonance $\omega_0 = -\gamma_{\text{n}} B_0$ and determining the resulting spin polarization, where $\gamma_{\text{n}}$ is the gyromagnetic ratio of the neutron. Alternatively, the frequency of the resonance spin flip coils is kept constant and only their relative phase is scanned. This so-called phase-scan has the advantage that one performs all measurements at the resonance frequency, i.e.\ $\omega = \omega_0$ \cite{Piegsa16}.
An additional spin-dependent interaction applied for the time $T$ between the two  $\pi/2$-pulses induces a corresponding phase-shift $\Delta \varphi = \Delta \omega T$ of the oscillation pattern. Here, $\hbar \Delta \omega$ corresponds to a Zeeman-like energy splitting of the spin states due to the investigated interaction. In the well-established neutron EDM measurement scheme, the neutron spins precess freely in simultaneously applied static magnetic and electric fields. By performing two successive measurements, one with the fields oriented parallel and a second with the fields oriented anti-parallel, one intends to determine the subtle phase-shift due to an interaction of the electric dipole moment $d_{\text{n}}$ with the electric field $E$. 
\begin{figure}
	\centering
		\includegraphics[width=0.48\textwidth]{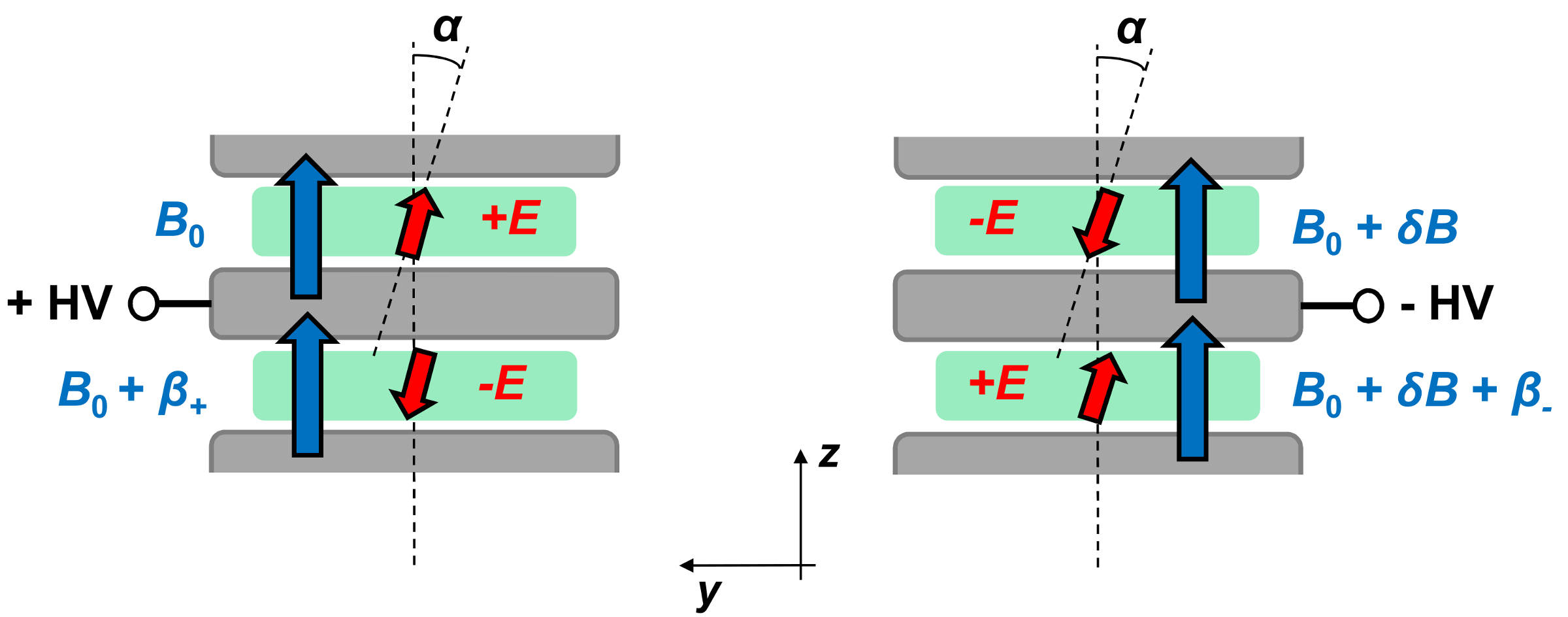}
	\caption{Scheme of the high voltage electrodes for the two different polarity settings (left: positive high-voltage and right: negative high-voltage). The two neutron beams, with their cross sections indicated by the green rectangles, are traveling between the electrodes in the $x$-direction. The angle $\alpha$ describes a potential misalignment between the electric and magnetic field. Here, $\delta B$ represents a global drift of the magnetic field between the two measurements, while $\beta_+$ and $\beta_-$ describe a systematic change of the magnetic field gradient.}
	\label{fig:Fig1_Electroden}
\end{figure}  \\
In Fig.\ \ref{fig:Fig1_Electroden} a schematic sectional drawing of the experimental setup is presented. Two separate neutron beams are traveling between three electrodes along the positive $x$-direction. Their spins precess in the $x$-$y$-plane perpendicular to the applied magnetic field $B_0$. The upper and lower electrodes are set to ground potential while the middle electrode is connected to a high-voltage source, thus providing opposite electric field directions in the two beams. By changing the polarity of the high-voltage the electric fields are inverted. However, in both cases a potential misalignment between the electric and magnetic field causes a first order $v \times E$-effect sensed by the neutrons. For small $\alpha$, i.e.\ $\sin \alpha \approx \alpha$ and $\cos \alpha \approx 1$, the corresponding phase-shifts in the upper (1) and lower (2) beam due to the electric field reversal are given by:
\begin{align}
  % \Delta \varphi_{1}  =  \left[ \gamma_{\text{n}} \delta B  - \frac{4 d_{\text{n}} E}{\hbar} \right]\cdot T - \frac{2 \gamma_{\text{n}} L E}{c^2} \sin \alpha \\
	% \Delta \varphi_{2} = \left[ \gamma_{\text{n}} \left( \delta B + \beta_- - \beta_+ \right) + \frac{4 d_{\text{n}} E}{\hbar} \right]\cdot T + \frac{2 \gamma_{\text{n}} L E}{c^2} \sin \alpha
	 \Delta \varphi_{1}  =  \left[ \gamma_{\text{n}} \delta B  - \frac{4 d_{\text{n}} E}{\hbar} \right]\cdot T - \frac{2 \alpha \gamma_{\text{n}} L E}{c^2}  \\
	 \Delta \varphi_{2} = \left[ \gamma_{\text{n}} \left( \delta B + \beta_- - \beta_+ \right) + \frac{4 d_{\text{n}} E}{\hbar} \right]\cdot T + \frac{2 \alpha \gamma_{\text{n}} L E}{c^2}
\label{eq:phaseshift1+2}
\end{align}
where $T$ is not the neutron time-of-flight from the source to the detector, but the neutron interaction time with the electric field. It is linked to the length of the electrodes $L$ via $v = L/T$. Thus, the total relative phase-shift of the two beams is:
\begin{equation}
  % \Delta \varphi_{2}  - \Delta \varphi_{1}  =  \left[ \gamma_{\text{n}} \left(  \beta_- - \beta_+ \right) + \frac{8 d_{\text{n}} E}{\hbar} \right]\cdot T + \frac{4 \gamma_{\text{n}} L E}{c^2} \sin \alpha
	\Delta \varphi_{2}  - \Delta \varphi_{1}  =  \left[ \gamma_{\text{n}} \left(  \beta_- - \beta_+ \right) + \frac{8 d_{\text{n}} E}{\hbar} \right]\cdot T + \frac{4 \alpha \gamma_{\text{n}} L E}{c^2} 
\label{eq:totphase}
\end{equation}
Note, the important advantage of using two beams is that possible global drifts $\delta B$ of the external magnetic field occurring during the polarity change cancel. 
This is equivalent to ultracold neutron EDM experiments with two separate precession chambers.
However, systematic gradient drifts represented by the magnetic fields $\beta_+$ and $\beta_-$ can still mimic a false neutron EDM signal:
\begin{figure}
	\centering
		\includegraphics[width=0.48\textwidth]{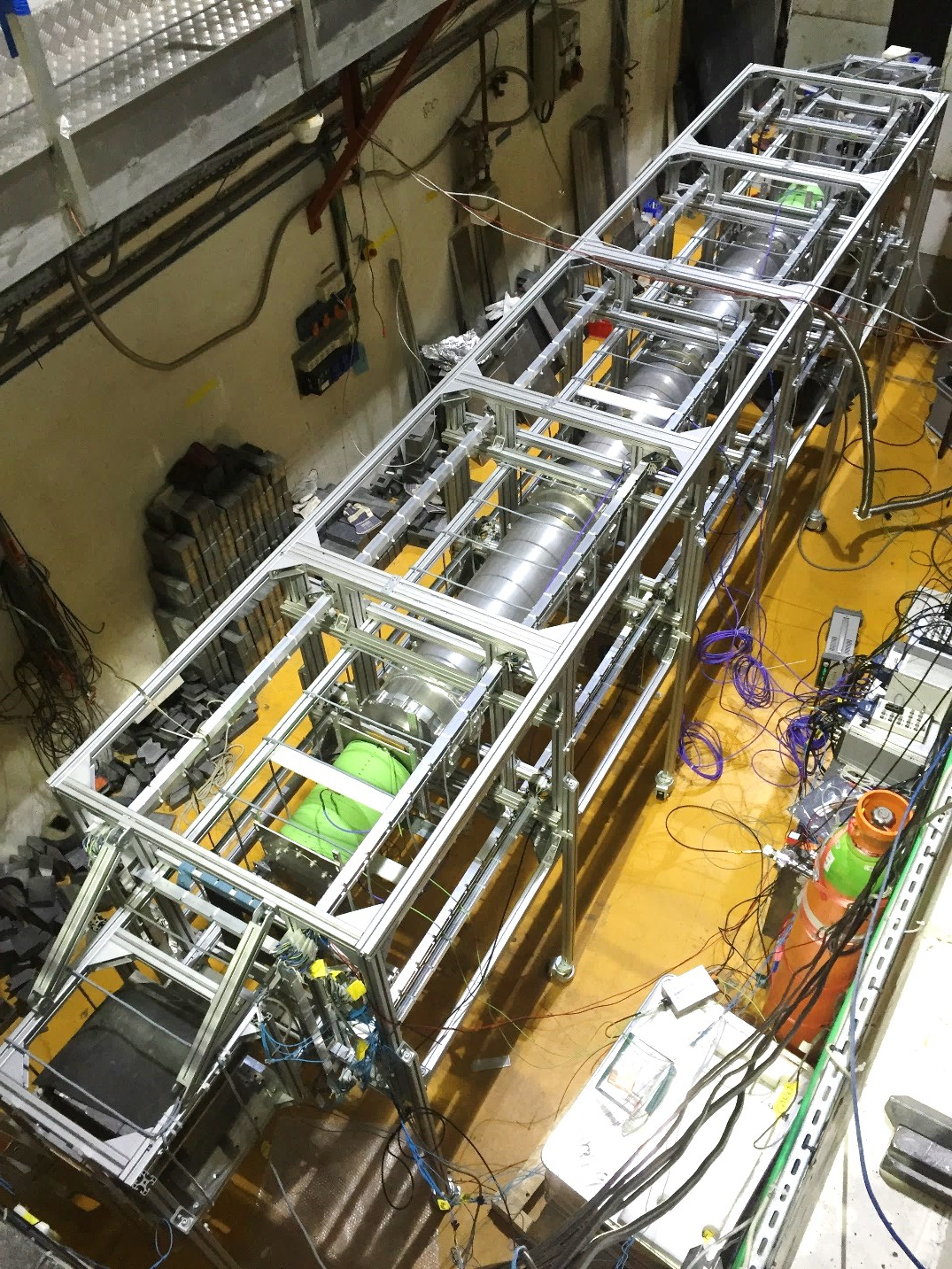}
	\caption{Installation of the Beam EDM prototype experiment at the PF1b beam line at the Institut Laue-Langevin. The total length of the setup is approximately seven meters. The neutron beam exits the wall at the experimental site in the lower left corner of the image. The neutron detector is located at the upper right corner of the image. The green coils are neutron resonance spin flip coils separated by 4~m. An approximately 3.4~m long aluminum vacuum pipe is installed between the coils.}
	\label{fig:Setup}
\end{figure} 
\begin{equation}
    d_{\text{n,false}}  =  \frac{\hbar \gamma_{\text{n}} \left( \beta_- - \beta_+ \right)}{ 8E}
\label{eq:flaseedm}
\end{equation}
For instance, to keep this false neutron EDM signal below $10^{-27}$~e~cm the systematic gradient difference correlated with the electric field reversal needs to be smaller than 10~fT at an electric field of 100~kV/cm.
A more general discussion of the scheme, including asymmetric electric field magnitudes, a higher order expansion, and other potential systematic false effects, is addressed in Ref.\ \cite{Piegsa13}.
Moreover, from Eq.\ (\ref{eq:totphase}) one can directly see that a time-of-flight measurement allows distinguishing between the interaction time dependent EDM effect (slope of the linear curve) and the interaction time independent $v \times E$-effect (offset).\\
The sensitivity of the neutron EDM beam experiment is given by the statistical uncertainty (standard deviation):
\begin{figure*}
	\centering
		\includegraphics[width=0.65\textwidth]{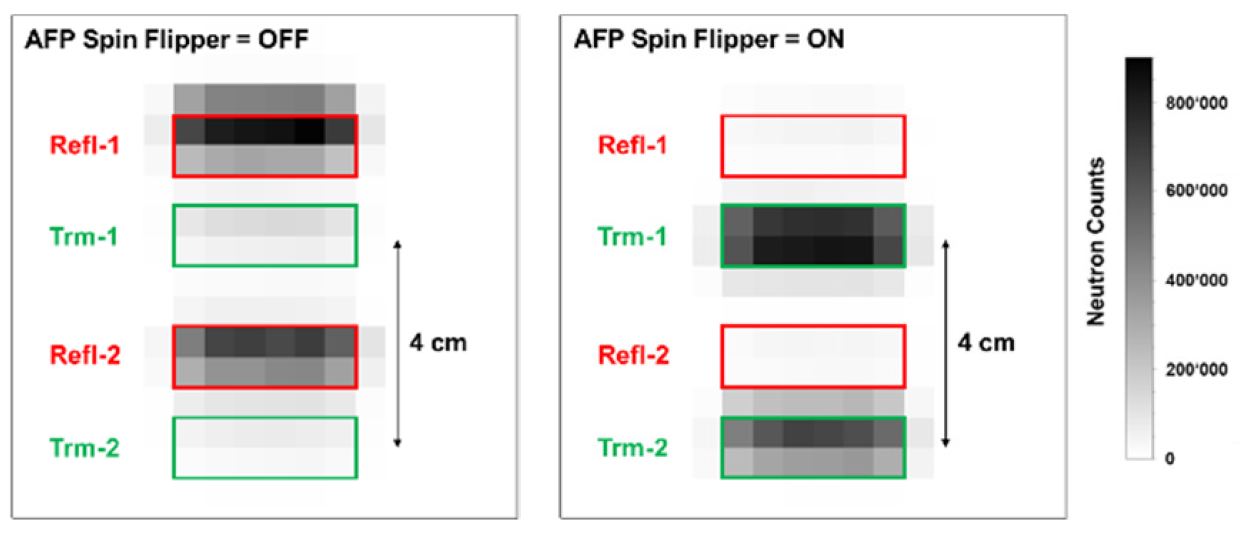}
	\caption{Detector counts in 10~s at a wavelength of 0.48~nm for the two adiabatic spin flip device states (off and on). The detector has an active area of $100 \times 100$~mm$^2$ divided into $16 \times 16$~pixels. The two beams cause four beam spots behind the spin analyzer – reflected (red box) and transmitted (green box) beam of the upper and lower beam, respectively. }
	\label{fig:Detector}
\end{figure*}
\begin{equation}
   \sigma (d_{\text{n}})  \approx  \frac{2 \hbar}{ \eta \tau E \sqrt{N}}
\label{eq:sensitivity}
\end{equation}
where $N$ is the total number of detected neutrons and $0 \leq \eta \leq 1$ is the visibility of the Ramsey fringe pattern. 
The statistical uncertainty is about a factor four larger than usually stated for UCN experiments, which arises from the necessity of performing the linear fit \cite{Golub72}. 
The basis of the neutron EDM measurement with a pulsed beam is to compensate the approximately 1000 times shorter interaction time by means of a larger electric field by a factor of about 10 and of higher statistics by a factor of about $10^6$ compared to state-of-the-art UCN experiments \cite{Pendle15}. 
% ... example ILL/RAL/Sussex ... The neutron count rate in the latest UCN experiment was approximately 60 s−1, as an average of 14000 UCN were detected per 240-s-long measurement cycle 
In Eq.\ (\ref{eq:sensitivity}) $\tau$ represents the interaction time interval over which the linear fit is performed:
\begin{equation}
%   \tau = T \left(\lambda_{\text{max}}\right) - T \left(\lambda_{\text{min}}\right)  = \frac{m_{\text{n}} L}{2 \pi \hbar} \left(\lambda_{\text{max}} - \lambda_{\text{min}}\right)
	\tau  = \frac{m_{\text{n}} L}{2 \pi \hbar} \left(\lambda_{\text{max}} - \lambda_{\text{min}}\right)
\label{eq:tautau}
\end{equation}
where $m_{\text{n}}$ is the mass of the neutron and $\lambda_{\text{max}}$ and $\lambda_{\text{min}}$ are the maximum and minimum neutron de Broglie wavelength of the employed neutron energy spectrum used in the measurement, respectively.
From this it becomes clear that the choice of the neutron spectrum has to be optimized, with respect to its flux, in order to achieve highest precision.
%\footnote{Note, Eq.\ (\ref{eq:sensitivity}) assumes equal neutron statistics at each velocity, i.e.\ it does not yet take the actual shape of energy spectrum into account.} 
To avoid overlap between pulses $\tau$ is limited by the pulse frequency $f_{\text{p}}$ of the neutron source:
\begin{equation}
  \tau \leq \frac{L}{f_{\text{p}} L_{\text{SD}}}
\label{eq:taulimit}
\end{equation}
where $L_{\text{SD}}$ is the total source-to-detector distance. Hence, using Eq.\ (\ref{eq:tautau}) this leads to a maximum neutron wavelength band:
\begin{equation}
   \lambda_{\text{max}} - \lambda_{\text{min}}  \leq \frac{2 \pi \hbar}{m_{\text{n}}  f_{\text{p}}  L_{\text{SD}}}
\label{eq:lambdalimit}
\end{equation}
For instance, in case of the ESS with $f_{\text{p}} = 14$~Hz and an assumed total length $L_{\text{SD}}=75$~m this yields a band width of about 0.4~nm. 
A doubling of the band width can be achieved by skipping every second spallation pulse by means of an additional chopper. Obviously, this comes together with a reduction in statistics, however, it might still be advantageous in view of the total sensitivity and in the investigation of systematic effects. 
The optimization of the stated neutron wavelength spectrum is currently under investigation. \\
Finally, assuming a visibility $\eta=0.75$, an effective interaction time $\tau=50$~ms (with a pulse frequency 14~Hz, $L=50$~m, and $L_{\text{SD}}=75$~m), an electric field  $E=100$~kV/cm, and a count rate $\dot{N} = 400$~MHz at the ESS, this yields a neutron EDM sensitivity of approximately $5 \times 10^{-26}$~e~cm in one day of data taking.\footnote{In a comparable experiment with a similar configuration fields of 80-130~kV/cm were reached \cite{Dress77}.}$^,$\footnote{The count rate was estimated in a previous publication assuming a comparable time-integrated neutron flux/brightness at ESS and at ILL \cite{Abele06}.}
% Hence, in 400 days of data taking the present best neutron EDM limit can be improved by about an order of magnitude.
Note, this differs from the previously stated value mainly due to the newly introduced factor four in Eq.\ (\ref{eq:sensitivity}) accounting for the linear fit \cite{Piegsa13}.

\section{Proof-of-principle apparatus}

\begin{figure}
	\centering
		\includegraphics[width=0.48\textwidth]{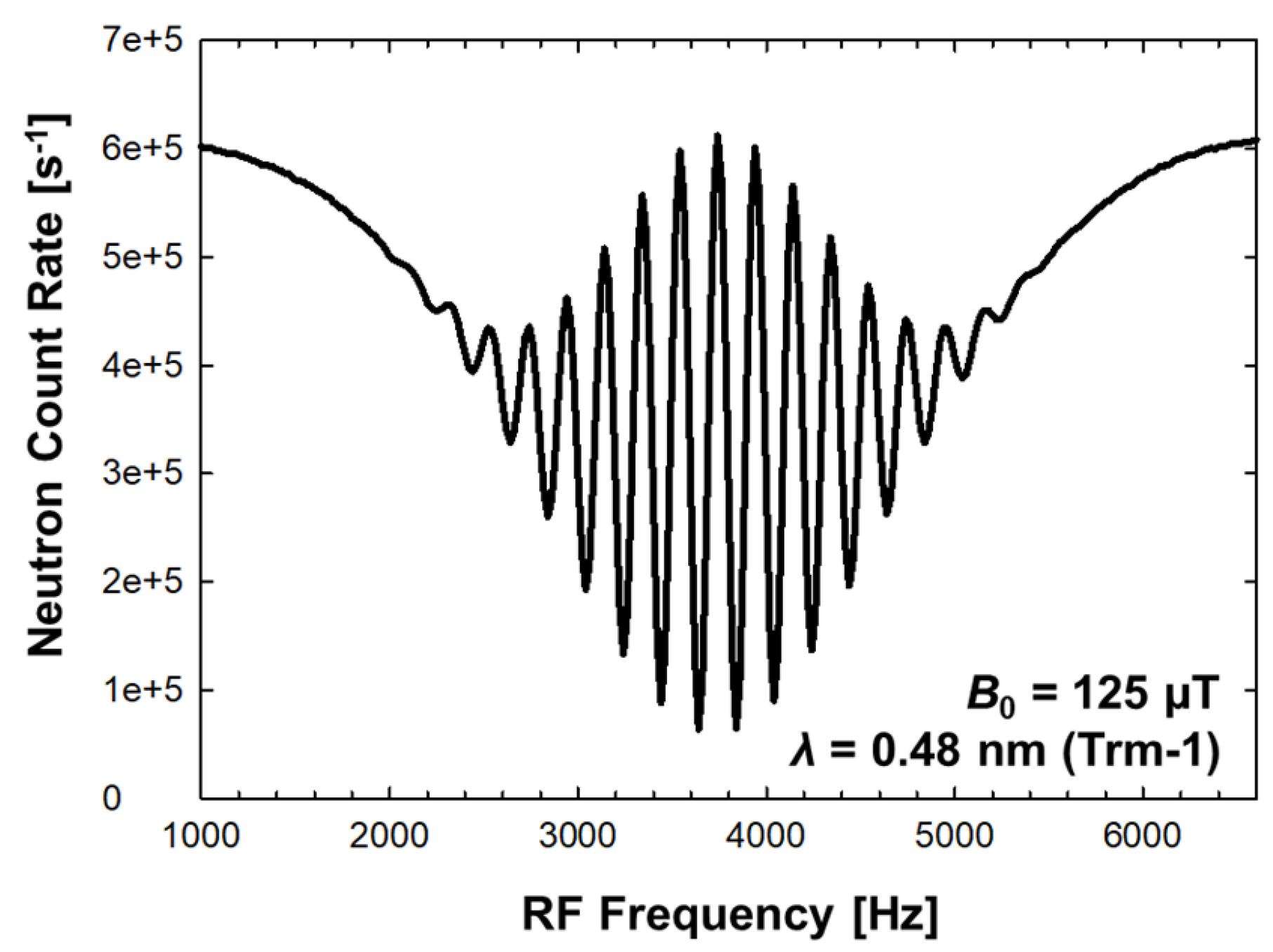}
	\caption{Ramsey resonance signal of one of the four detector beam spots (upper beam, transmission) at a wavelength selector setting of 0.48~nm with a wavelength spread of $\Delta \lambda / \lambda \approx 10$\%. The presented result is obtained by scanning the frequency of the spin flip coils close to the neutron Larmor resonance frequency. The entire curve was measured within approximately one hour (10~s per point).}
	\label{fig:Ramsey}
\end{figure}

Currently, we are setting up and characterizing a prototype Ramsey apparatus. 
In Fig.\ \ref{fig:Setup} a photograph of the setup during a beam time at PF1b at ILL is presented \cite{Abele06}.
%... The main aim of the three-weeks-long beam time at PF1b was to perform characterization measurements of the cold neutron precision Ramsey spectrometer presented in Fig. 1. The apparatus has been fully developed in the laboratories at the University of Bern. 
It consists of a non-magnetic aluminum structure which supports all elements of the Ramsey apparatus (collimator, apertures, spin flip coils, vacuum beam pipe, spin analyzer, and neutron detector) as well as all magnetic field coils. 
The structure consists of modular $1 \times 1 \times 1$~m$^3$ cubes which in a later stage can be covered with a passive magnetic shielding of mu-metal. \\
During the beam time at ILL a multitude of investigations concerning the performance of the apparatus were conducted. 
To perform systematic studies an adiabatic spin flip device and a neutron velocity selector were installed in the bunker of PF1b. 
Further, the primary beam was split into two parallel well-collimated beams using apertures (upper and lower beam) each with a size of 1 (height) $\times$ 4 (width) cm$^2$, with a center-to-center distance of 4~cm. 
In front of the position sensitive detector which is capable of high-rates \cite{Cascade}, the two beams were spin analyzed by means of two sets of spin polarizer blades (silicon wafers coated with $m=5$ FeSi multi-layers, one spin state gets reflected the other transmitted). This results in four well-separated beam spots on the detector – a reflected and a transmitted spot for each beam. An image of a typical detector intensity pattern is presented in Fig.\ \ref{fig:Detector}. This four-beam-pattern allows for normalization of the intensity to compensate for slight fluctuations of the neutron output of the cold source. 
%During this beam time, it was not planned for the high-voltage electrodes to be installed. 
%Hence, the two beams will only in later experiments pass between three electrodes – two ground electrodes and a central high-voltage electrode. Hence, the upper and lower beam will experience opposite electric fields and thus can be used to compensate for common-mode noise and drifts. 
The static magnetic field in our setup is oriented vertically and typically operates at a field of about $125$~$\mu$T. The field is actively stabilized to the nT-level using a feedback signal from an array of fluxgates.
In Fig.\ \ref{fig:Ramsey} a first Ramsey pattern is presented which was obtained by scanning the frequency of our resonance spin flip coils. 
% In the next beam times, we plan to employ and characterize the high-voltage electrodes and the aforementioned passive magnetic shield. Eventually, this will allow us to take first real neutron EDM data with our apparatus.

\section{Conclusion and outlook}

We have presented the current status of the neutron Beam EDM project. 
So far, we have conducted several first test measurements in beam times at PSI and ILL. 
In the coming years, we will continue performing further proof-of-principle studies investigating the measurement technique and potential systematic effects. The next steps will consist of integrating a high-voltage electrode system, using a pulsed/chopped neutron beam, and designing the aforementioned passive magnetic shield.
Eventually, this will allow us to take first real neutron EDM data with our apparatus.
With this setup, we aim for a considerable improvement of the present best neutron beam EDM limit given in Ref.\ \cite{Dress77}.
Ultimately, the full-scale experiment shall be realized at the upcoming ESS.

% investigating the feasibility of the experiment at the ESS Continue with proof-of-principle

% implement a pulsed beam using choppers ... modulation of $B_1$ rf signal
% investigate systematic effects
% aim for a considerable improvement compared to the current best beam EDM measurement \cite{Dress77} ... \\

% More technical ... modulation of $B_1$ rf signal 

% ... needs more details ... what assumptions ... and $\tau$ corresponds to covered wavelength band ... more wide is better ... ... what happens if only measure at two wavelength points ... and equally over an entire band with equidistant points ... maybe explain in these steps ...

\section*{Acknowledgments}
This work was supported via the European Research Council under the ERC Grant Agreement no.\ 715031 and via the Swiss National Science Foundation under the grant no.\ PP00P2-163663.

% \newpage 


\begin{thebibliography}{}

\bibitem{Pendle15}{}     J.M. Pendlebury et al., Phys. Rev. D \textbf{92}, 092003 (2015)

\bibitem{Baker11}{}   	 C.A. Baker et al., Phys. Proc. \textbf{17}, 159 (2011) 
\bibitem{Abel18}{}       C. Abel et al., these proceedings, arxiv: 1811.02340 
%\bibitem{Serebrov14}{}  A.P. Serebrov et al., JETP Lett. \textbf{99}, 4 (2014) 
\bibitem{Serebrov15}{}   A.P. Serebrov et~al., Phys. Rev. C \textbf{92}, 055501 (2015)
\bibitem{Ito07}{}   	   T.M. Ito, J. Phys. Conf. Ser. \textbf{69}, 012037 (2007) 
\bibitem{Tsentalovich2014}{} E.P. Tsentalovich, Physics of Particles and Nuclei {\bf 45}, 249 (2014)
\bibitem{Lamoreaux09}{}  S.K. Lamoreaux and R. Golub,  J. Phys. G: Nucl. Part. Phys. \textbf{36}, 104002 (2009) 
\bibitem{Altarev12}{}    I.S. Altarev et al.,  Nuovo Cimento C \textbf{35}, 122 (2012) 
\bibitem{Picker17}{}	   R. Picker, JPS Conf. Proc. \textbf{13}, 010005 (2017)
\bibitem{Masuda2012}{}   Y. Masuda et al., Phys. Lett. A {\bf 376} 1347 (2012)






\bibitem{Dress77}{}      W.B. Dress, P.D. Miller, J.M. Pendlebury, P. Perrin and N.F. Ramsey, Phys. Rev. D \textbf{15}, 9 (1977)
\bibitem{Piegsa13}{}     F.M. Piegsa, Phys. Rev. C \textbf{88}, 045502 (2013)
\bibitem{Klinkby16}{}    E. Klinkby and T. Soldner, J. Phys. Conf. Ser. \textbf{746}, 012051 (2016)
\bibitem{Soldner18}{}    T. Soldner et al., these proceedings, arxiv: 1811.11692


\bibitem{Ramsey49}{}     N.F. Ramsey, Phys. Rev. \textbf{76}, 996 (1949)
\bibitem{Ramsey50}{}     N.F. Ramsey, Phys. Rev. \textbf{78}, 695 (1950)

\bibitem{Piegsa08}{}     F.M. Piegsa et al.,  Nucl. Instrum. and Meth. A \textbf{589}, 318 (2008) 
\bibitem{Piegsa09}{}     F.M. Piegsa et al., Phys. Rev. Lett. \textbf{102}, 145501 (2009)
\bibitem{Piegsa12}{}     F.M. Piegsa and G. Pignol, Phys. Rev. Lett. \textbf{108}, 181801 (2012)
\bibitem{Piegsa14}{}	   F.M. Piegsa, Physics Procedia \textbf{51}, 59 (2014)


\bibitem{Piegsa16}{} 	   F.M. Piegsa, P. Hautle and C. Schanzer, Phys. Rev. C \textbf{93}, 045501 (2016)
\bibitem{Golub72}{}      R. Golub and J.M. Pendelbury, Contemp. Phys. \textbf{13}, 519 (1972)

\bibitem{Abele06}{}	     H. Abele et al., Nucl. Instrum. and Meth. A \textbf{562}, 407 (2006) 

\bibitem{Cascade}{}      Cascade Detector Technologies, www.n-cdt.com

%Journal Author, Journal \textbf{Volume}, page numbers (year)
%Book Author, \textit{Book title} (Publisher, place, year) page numbers
% etc

\end{thebibliography}
\end{document}